\documentclass[preprint,prl,final]{revtex4}

\usepackage[T1]{fontenc}
\usepackage{graphicx}
\usepackage{amsfonts}
\usepackage{amssymb}
\usepackage{amsmath}
\usepackage{color}

\begin{document}

\title{Thickness dependence of the degree of spin polarization of the electrical current in permalloy thin films}

\author{Mohammad Haidar}
\email{mohammad.haidar@hotmail.fr}
\author{Matthieu Bailleul}
\email{matthieu.bailleul@ipcms.unistra.fr}
\affiliation{IPCMS and NIE, UMR 7504 CNRS-Universit\'{e} de Strasbourg, 23 rue du Loess, BP 43, 67034 Strasbourg Cedex 2, France}


\date{\today}

\begin{abstract}

Spin-polarized electrical transport is investigated in $ Al_{2}O_{3}/Ni_{80}Fe_{20}/Al_{2}O_{3}$ thin films for permalloy thickness between 6 and 20nm. The degree of spin-polarization of the current flowing in the plane of the film is measured through the current induced spin wave Doppler shift. We find that it decreases as the film thickness decreases, from 0.72 at 20nm to 0.46 at 6nm. This decrease is attributed to a spin depolarization induced by the film surfaces. A model is proposed which takes into account the contributions of the different sources of electron scattering (alloy disorder, phonons, thermal magnons, grain boundaries, film surfaces) to the measured spin-dependent resistivities.


\end{abstract}
\maketitle

The flow of a spin polarized current through a non-uniform magnetization distribution is able to transfer angular momentum to the local magnetization \cite{Berger1996,Slonczewski1996}. This spin transfer torque is now used in a number of spintronic devices, in particular those involving current induced domain-wall motion \cite{Parkin2008}. The degree of spin polarization of the electrical current, defined as the contrast between the currents carried by the majority and minority electrons ($P=\frac{J_{\uparrow}- J_{\downarrow}}{J_{\uparrow}+ J_{\downarrow}}$), is an essential parameter controlling the performance of these devices. In the early days of spintronics, the contribution of the impurities to $P$ was extracted indirectly from low temperature resistance measurements on bulk dilute alloys \cite{FertCampbell}. However, these estimates are not directly applicable to the very thin ferromagnetic films used for spintronic devices in which additional sources of electron scattering contribute to $P$. In particular, it is expected that film surfaces play an important role as soon as the film thickness becomes comparable to the bulk electron mean fee path \cite{Sondheimer1952}. Although some information could be extracted indirectly from giant magnetoresistance measurements \cite{Dieny1992,gurney1993}, there is no accurate study of the influence of film surfaces on the degree of spin polarization. In order to address this question, we have measured the film thickness dependence of $P$, resorting to the technique of the Current Induced Spin Wave Doppler Shift (CISWDS) which gives a direct access to  it \cite{Vlaminck2008}. When a DC current $I$ passes along a ferromagnetic metal strip of width $w$ and thickness $t$, the spin transfer torque modifies the propagation of a spin-wave of wave-vector $k$ in the form of a Doppler shift $\Delta f_{Dop}$ of the frequency $f$ of the spin-wave which writes:
\begin{equation}\label{eq:1}
\Delta f_{Dop}=-\frac{\mu_{B}}{2\pi e} \frac{P}{M_{s}} \frac{I}{t w} k,
\end{equation}
where $\mu_{B}$ is the Bohr magneton, $e$ is the magnitude of the electron charge and $M_{s}$ is the saturation magnetization. After our initial measurement on a permalloy film at room temperature \cite{Vlaminck2008}, the CISWDS technique has been extended to low temperature \cite{Zhu2010}, to other materials \cite{Zhu2011,Thomas2011} and to time-domain measurements \cite{Sekiguchi2012}. In this letter, we investigate the CISWDS as a function of the film thickness in a very common system of spintronics, namely polycrystalline permalloy films of thickness between 6 and 20nm. The measured thickness dependence of $P$ is interpreted within a modified two-current model accounting for all the electron scattering processes relevant for these films.

\begin{figure}[!htp]
\begin{center}
\includegraphics[scale=0.4]{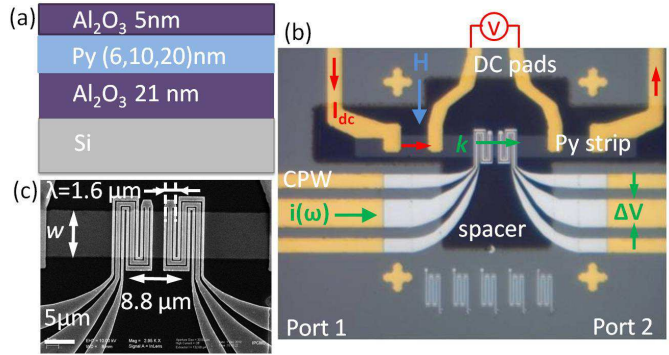}
\caption{(a) Film stack. (b) Optical microscope image of the CISWDS device. One recognizes the strip ion milled from this stack, the four DC pads and the two coplanar waveguides (Ti 10nm/Au 60nm), the insulating spacer ($SiO_{2}$ 80nm) and the two spin-wave antennae (Ti 10nm/Al 120nm). The conventions used in the text for the directions of positive k, I and H are shown. (c) Scanning electron microscope image showing the meander shape of the antennas.}\label{fig:1}
\end{center}
\end{figure}

\begin{figure}[!htp]
\begin{center}
\includegraphics[width=0.35\textwidth]{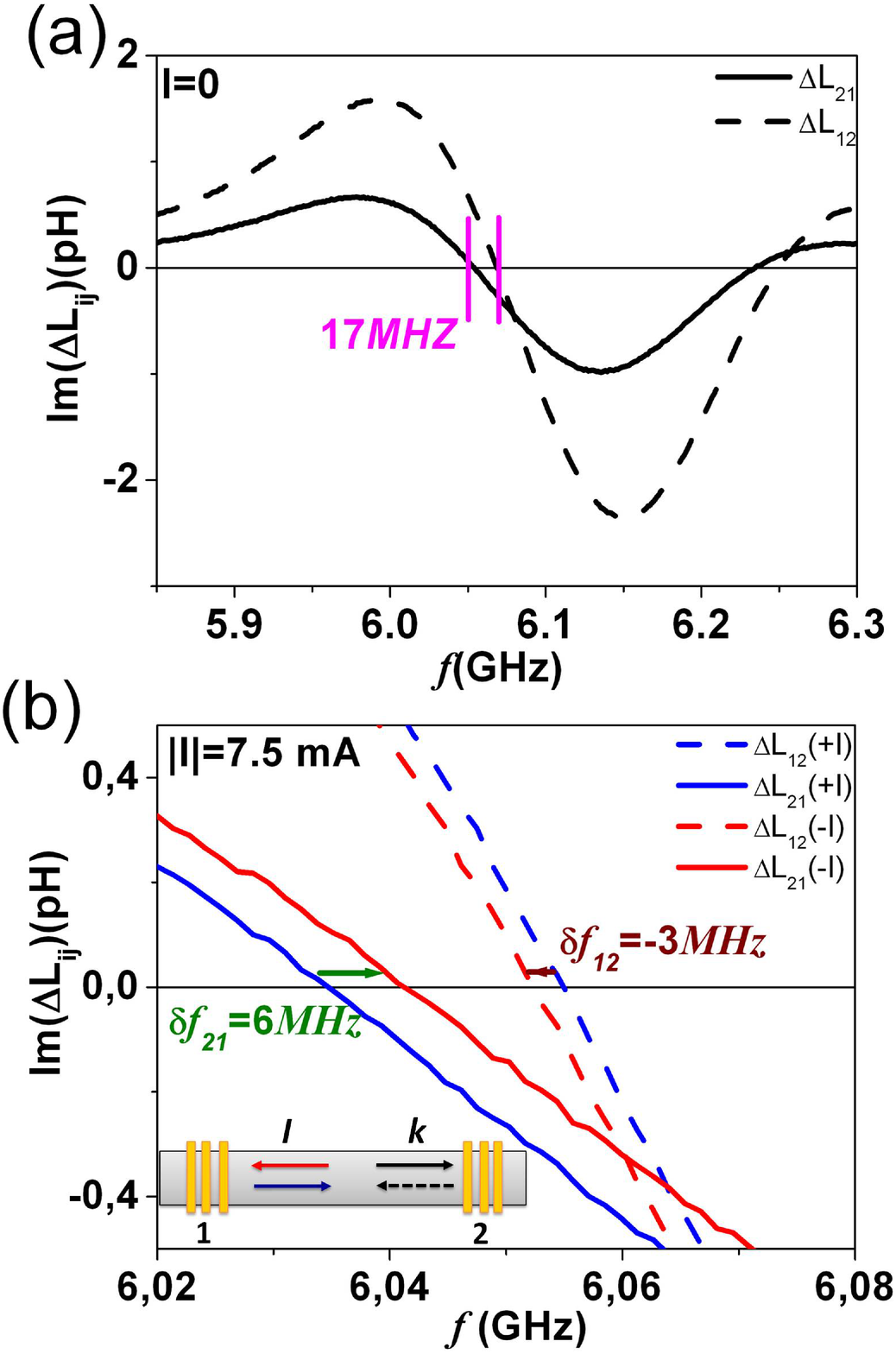}
\caption{(a) Imaginary part of the mutual inductance signals $\Delta L_{12}$ and $\Delta L_{21}$ measured on the 10nm device under a field of 28mT and in the absence of DC current. (b) Same for a DC current of magnitude $|I|= 7.5 mA$. The frequency range for which the curves intersect the $x$ axis has been zoomed in for clarity.}\label{fig:2}
\end{center}
\end{figure}

Permalloy (Py) films of thickness 6, 10 and 20nm sandwiched by $Al_{2}O_{3}$ layers were grown by magnetron sputtering on silicon substrates [Fig.~\ref{fig:1}(a)]. For each film thickness, we fabricated CISWDS devices which comprises a ferromagnetic strip of width $w= 8\mu m$, a pair of antennae with meander shape allowing the excitation and the detection of spin waves with a wavevector $k= 3.86 \mu m^{-1}$, and four DC pads serving to pass the DC current $I$ through the ferromagnetic strip and to measure its resistance [Fig.~\ref{fig:1}(b)]. The propagating spin wave spectroscopy measurements are performed as described in \cite{Vlaminck2010,haidarPhD,Vincentthesis}. In the present work, the external magnetic field $\mu_0H_{0}=28$mT is applied  in the plane of the film perpendicular to the strip so that the spin waves propagate in the so-called magnetostatic surface wave (MSSW) configuration \cite{Stancil2009}. Compared to the magnetostatic forward volume waves (MSFVW) used in our first report \cite{Vlaminck2008}, MSSWs have the advantage of providing propagating spin wave signals of higher amplitudes and of requiring a lower external field. This is at the price of an increased complexity for extracting the CISWDS which has to be separated from artefactual contributions (see explanations below).
Let us first describe the signals measured in the absence of DC current. The solid curve in Fig.~\ref{fig:2}(a) shows the imaginary part of the mutual inductance $\Delta L_{21}$ \footnote{The $\Delta L_{ij}$s are obtained by substracting a reference signal registered at 190mT (see measurement procedure in \cite{Vlaminck2010}).} which corresponds to spin waves propagating from antenna 1 to antenna 2 [$+k$ on Fig. \ref{fig:1}(b)]. The dashed curve shows the mutual inductance $\Delta L_{12}$ (spin waves propagating from antenna 2 to antenna 1, $-k$). One sees immediately that the amplitude for $+k$ is about three times larger than that for $-k$. This amplitude non reciprocity is a specific feature of MSSWs which has already been observed both in garnet \cite{Schneider2008} and permalloy \cite{bailleul2003,Sekiguchi2010} films. It originates from the fact that the polarization of the dynamical magnetization matches better the polarization of the antenna field for one direc-tion than for the other \cite{Schneider2008}. We also observe in Fig.~\ref{fig:2}(a) that the $+k$ signal is shifted 17MHz higher in frequency than the $-k$ one. This frequency non-reciprocity is attributed to the combination between the wave localization non-reciprocity (MSSWs have more amplitude at one surface than on the other, this localization depends on the sign of $k$ and $H$) and an asymmetry of the film \cite{Amiri2007}. As expected, both amplitude and frequency non-reciprocities are reversed when the external field is reversed (not shown). Due to these non-reciprocal features, the method we have used previously to extract the CISWDS \cite{Vlaminck2008}, i.e. compare the frequency of counter propagating spin-waves at a given current, cannot be used. Instead, we have chosen to compare measurements taken for opposite currents.

Fig.~\ref{fig:2}(b) shows the current-induced frequency shifts at $|I|=7.5mA$: for $+k$ the $-I$ curve lies at a higher frequency than the $+I$ one ($\delta f_{21}=f_{21}(-I)-f_{21}(+I)=+6MHz$), whereas for $-k$ the $-I$ curve lies at a lower frequency than the $+I$ one ($\delta f_{12}=f_{12}(-I)-f_{12}(+I)=-3MHz$). The signs of the observed shifts are in agreement with the CISWDS expected for $P>0$, namely an increase of the frequency when the spin wave propagates along the electron flow, i.e. against the current. However, the magnitude of  $\delta f_{12}$ and  $\delta f_{21}$ are different, in contrast to what is expected for a pure Doppler effect. This indicates that another phenomenon, with a different symmetry with respect to $k$, also contributes to the observed current-induced shifts. We believe that the Oersted field generated by the DC current is the origin of this additional contribution: due to an asymmetry across the film thickness, the Oersted field contribution to the spin-wave frequency does not average out strictly to zero. This contribution does not change sign when $k$ is reversed. As expected, it does change sign when $H$ is reversed (not shown). Finally, the current-induced spin-wave Doppler shift is extracted using the following relation:
\begin{equation}
\Delta f_{Dop}=\frac{\delta f_{21}- \delta f_{12}}{4},
\end{equation}
where the $\delta f_{ij}$ are the current-induced frequency shifts defined above~\footnote{For extracting the current-induced shifts with a high precision, we use $\delta f_{ij}=f_{per}Im(R_{ij})/2 \pi$, where $f_{per}$ is the period of the oscillating mutual-inductance signal and $R_{ij}=(\Delta L_{ij}(-I)- \Delta L_{ij}(+I))/ (\Delta L_{ij}(-I)+ \Delta L_{ij}(+I))/$ is the contrast between the $+I$ and $-I$ complex signals \cite{haidarPhD}.}. This procedure allows one to separate the CISWDS from the unwanted contributions associated to the MSSW non-reciprocity and to the Oersted field. It avoids the reproducibility issues associated with the reversal of the external field \cite{Zhu2010} and gives an overall precision of the order of $100kHz$.

\begin{figure}[!htp]
\begin{center}
\includegraphics[width=0.35\textwidth]{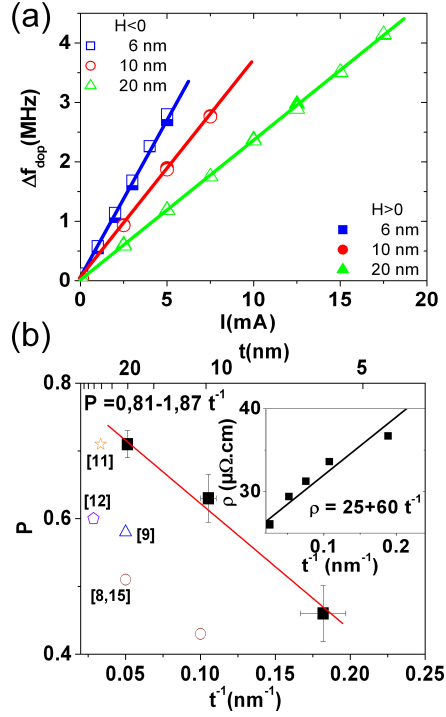}
\caption{(a) Doppler frequency shift as a function of the DC current for 6, 10, and 20nm thick films under an applied field of $\pm 28mT$. (b) Degree of spin-polarization of the electrical current as a function of the inverse of the film thickness. The inset shows the measured resistivity of each film. The colored points show published values.}\label{fig:3}
\end{center}
\end{figure}

Figure~\ref{fig:3}(a) shows the values of the current induced spin wave Doppler shift as a function of the DC current for different film thicknesses. Interestingly, the shifts measured for a negative field (open symbols) are identical to those obtained for a positive field (closed symbols), which confirms that our procedure is capable of eliminating the two artefacts described above. For each film thickness, one obtains a clear linear dependence. From the slopes, we extract the degree of spin-polarization of the electrical current $P$ using Eq.~(\ref{eq:1}) together with the width $w$ of the strip deduced from the SEM images and the product $\mu_{0}M_s t$ deduced from SQUID measurements. The obtained spin-polarization is plotted in Fig.~\ref{fig:3}(b) as a function of the inverse of the film thickness. One recognizes a strong decrease of the polarization as the thickness decreases, from $P=0.72$ at $t=20nm$ to $P=0.46$ at $t=6nm$. The $1/t$ fit displayed as a thin line on Fig.~\ref{fig:3}(b) indicates a bulk extrapolate $P=0.81$ and a critical thickness for which the polarization extrapolates to zero of $2.3nm$. This unexpectedly strong thickness dependence constitutes the main finding of the present paper. A natural way to interpret this result is to consider that the film surfaces tend to depolarize the electrical transport: as the film thickness decreases, the role of the film surfaces is enhanced, resulting in a lower polarization compared to bulk. In the remaining of the paper we give a quantitative explanation of this behavior.\\

\begin{figure}[!htp]
\begin{center}
\includegraphics[width=0.45\textwidth]{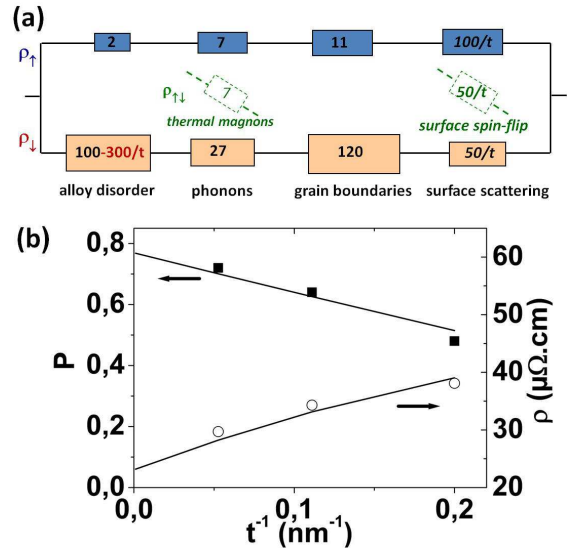}
\caption{(a) Model for the spin-polarized transport in our permalloy films. (b) Spin-polarization (left scale) and resistivity (right scale) versus the inverse of the film thickness. The measured values are shown as symbols and the results of the model are shown as lines.}\label{fig:4}
\end{center}
\end{figure}

To account for the polarization values of Fig.~\ref{fig:3}(b) together with the resistivity values measured on the same devices [inset in Fig.~\ref{fig:3}(b)], we propose the two-current model sketched in Fig.~\ref{fig:4}(a). The contributions of the different sources of electron scattering to the spin-dependent resistivities $\rho_{\uparrow}$ and  $\rho_{\downarrow}$ of our films, namely alloy disorder, phonons, grain boundaries and surface roughness, are assumed to sum up in series within each channel. Indeed, we believe the Mathiesen's rule to remain a reasonable approximation even for surface \cite{Sondheimer1952} and grain boundary \cite{Mayadas1970} scattering. A spin-flip resistivity $\rho_{\uparrow\downarrow}$ is also introduced to account for spin-mixing processes \cite{FertCampbell}. Within this model, the resistivity and the degree of spin-polarization of the current write respectively:
\begin{eqnarray}\label{eq:2}
\rho=\frac{\rho_{\uparrow}\rho_{\downarrow}+\rho_{\uparrow\downarrow}(\rho_{\downarrow}+\rho_{\uparrow})}
{\rho_{\uparrow}+\rho_{\downarrow}+4\rho_{\uparrow\downarrow}},\\
P=\frac{\rho_{\downarrow}\rho_{\uparrow}}{\rho_{\uparrow}+\rho_{\downarrow}+4\rho_{\uparrow\downarrow}}.
\end{eqnarray}
Let us now discuss the contribution of each scattering mechanism individually, starting with the bulk mechanisms. The contribution of the alloy disorder has been estimated both experimentally \cite{FertCampbell} and theoretically \cite{Mertig1993,Banhart1997}. It is known that Fe atoms in a Ni matrix act as strongly spin-dependent scattering centers \cite{Mijnarends2002}. Indeed, the majority electron local density of states (DOS) on Fe and Ni match very well ($\rho_{\uparrow}$= $2 \mu \Omega cm$) whereas the minority electrons DOS are very different ($\rho_{\downarrow}$= $100 \mu \Omega cm$). For the phonon contribution, we use the pure nickel room temperature estimate ($\rho_{\uparrow}$= $7 \mu \Omega cm$, $\rho_{\downarrow}$= $27 \mu \Omega cm$) extrapolated from the temperature dependence of the resistivity of dilute alloys \cite{FertCampbell}. For the spin-flip resistivity induced by thermal magnons, we use the value $\rho_{\uparrow\downarrow}$= $7 \mu \Omega cm$ deduced from temperature dependent CISWDS measurements \cite{Zhu2010}. Finally, we adjust the remaining bulk contributions i.e. that of the grain boundaries, to account for the bulk extrapolates of the measured resistivity ($\rho_{\uparrow}$= $25 \mu \Omega cm$) and polarization ($P=0.81$) we measure. We thus find that grain boundaries are responsible for significant contributions  ($\rho_{\uparrow}$= $11 \mu \Omega cm$, $\rho_{\downarrow}$= $120 \mu \Omega cm$) with a strong spin asymmetry. This asymmetry is ascribed to the fact that minority electrons, with their complex Fermi surface, are more sensitive to details of the local atomic ordering than majority ones \cite{Schulthess1997}.

Next, we discuss the contributions of the film surfaces to the electron scattering. Unless they are atomically smooth, surfaces are known to induce non-specular (diffuse) scattering which increases the film resistivity. This contribution can be approximated as $0.375(1-p)\rho_{bulk} \ell /t$ where $\rho_{bulk}$ is the bulk resistivity, $\ell$ is the bulk electronic mean free path and $p$ is the degree of specularity \cite{Sondheimer1952}. From current-in-plane giant magnetoresistance measurements, it was deduced that majority and minority electrons in permalloy have very different mean free paths of $\ell_{\uparrow}$=10 nm and $\ell_{\downarrow}$=0.5 nm respectively \cite{Dieny1992,gurney1993}. Combining those values with the bulk resistivities given above and assuming that scattering is completely diffuse ($p=0$), one obtains the values given in the rightmost boxes in Fig.~\ref{fig:4}(a), indicating that the spin-polarization for surface scattering is actually negative. This explains part of the decrease of spin-polarization at smaller film thicknesses. Another part of this decrease is certainly due to a change of stoechiometry due to the selective oxidation of Fe at the film surface. Indeed, X-ray photoelectron spectroscopy indicates that approximately 1.5 nm of iron oxide is formed between the permalloy layer and the top $Al_{2}O_{3}$ layer, whereas the Ni remains completely metallic. As a consequence, the remaining metal layer becomes depleted in Fe ($<C_{Fe}>=0.2-0.6/t_{[nm]}$, if one assumes that 0.75nm of iron is consumed by the oxidation). This modifies directly the alloy disorder contribution to the spin-dependent resistivities which, in good approximation, are linear with respect to $C_{Fe}$ \cite{Banhart1997}. While the decrease of the (already small) majority resistivity can be neglected, one gets a significant decrease of the minority resistivity $\rho_{\downarrow[\mu \Omega cm]}=500C_{Fe}=100-300/t_{[nm]}$ [bottom leftmost box in Fig.~\ref{fig:4}(a)]. Finally, we find it necessary to include a third ingredient, namely a thickness dependent spin-flip electron scattering [rightmost dashed box in Fig.~\ref{fig:4}(a)] to account for the decrease of spin-polarization at smaller film thicknesses. While the origin of this contribution is not clear at the moment, one is tempted to relate it to some spin-disorder at the interface between the permalloy and the iron oxide (presumably in an antiferromagnetic phase), similar to what has been observed in over-oxidized magnetic tunnel junctions \cite{Moodera1999}.

Fig.~\ref{fig:4}(b) illustrates the good agreement between the measured values of $P$ and $\rho$ and those obtained from Eq.~(\ref{eq:2}) using the model of Fig.~\ref{fig:4}(a).  Obviously, the modeling we propose contains many approximations and the numbers given in each box are to be taken with caution. However, we believe that the general picture is robust because significant discrepancies appear as soon as one removes one of the ingredients of the model. Before we conclude, let us remind these ingredients: (i) a robust strongly spin-polarized alloy disorder contribution (ii) a contribution from thermal excitations (phonons and thermal magnons) which tend to depolarize moderately the electrical current, (iii) a significant contribution from grain boundaries which is also quite strongly spin-polarized, and (iv) a contribution from diffuse surface electron scattering, which in our case has a small negative spin-polarization. We also identified two other effects, namely (v) depletion of iron and (vi) surface spin-flip scattering which we relate to surface oxidation and which might be specific to the film stack investigated.

To conclude, our study demonstrates the influence of surface effects on the spin-polarization of the current flowing in the plane of a thin ferromagnetic metal film. We believe such effects should be taken into account for spin transfer torque experiments carried out in such a geometry. More generally, we believe that our procedure, which combines CISWDS and resistivity measurements with a detailed two current model analysis, is the relevant one to identify the physical mechanisms governing the spin-dependent resistivities. From that point of view, systematic investigations of this kind could be very useful for the optimization of new film stacks for future STT applications.

Acknowledgements : We thank the STnano staff for their assistance in nanofabrication, M. Romeo and P. Bernhardt for the XPS measurements, and F. Gautier, Y. Henry and O. Bengone for stimulating discussions. This work was supported by the ANR (NanoSWITI, ANR-11-BS10-003).

\end{document}